\documentstyle[epsfig]{aprim10}
\input{epsf}
\newif\ifAMStwofonts

\ifoldfss
  \ifCUPmtlplainloaded \else
    \NewTextAlphabet{textbfit} {cmbxti10} {}
    \NewTextAlphabet{textbfss} {cmssbx10} {}
    \NewMathAlphabet{mathbfit} {cmbxti10} {} 
    \NewMathAlphabet{mathbfss} {cmssbx10} {} 
  \fi
  \ifAMStwofonts
    \ifCUPmtlplainloaded \else
      \NewSymbolFont{upmath} {eurm10}
      \NewSymbolFont{AMSa} {msam10}
      \NewMathSymbol{\upi}     {0}{upmath}{19}
      \NewMathSymbol{\umu}     {0}{upmath}{16}
      \NewMathSymbol{\upartial}{0}{upmath}{40}
      \NewMathSymbol{\leqslant}{3}{AMSa}{36}
      \NewMathSymbol{\geqslant}{3}{AMSa}{3E}

    \fi
  \fi
\fi 

\ifnfssone
  \newmathalphabet{\mathit}
  \addtoversion{normal}{\mathit}{cmr}{m}{it}
  \addtoversion{bold}{\mathit}{cmr}{bx}{it}
  \newmathalphabet{\mathbfit} 
  \addtoversion{normal}{\mathbfit}{cmr}{bx}{it}
  \addtoversion{bold}{\mathbfit}{cmr}{bx}{it}
  \newmathalphabet{\mathbfss} 
  \addtoversion{normal}{\mathbfss}{cmss}{bx}{n}
  \addtoversion{bold}{\mathbfss}{cmss}{bx}{n}
  \ifAMStwofonts
    \ifCUPmtlplainloaded \else
      \UseAMStwoboldmath
      \makeatletter
      \new@mathgroup\upmath@group
      \define@mathgroup\mv@normal\upmath@group{eur}{m}{n}
      \define@mathgroup\mv@bold\upmath@group{eur}{b}{n}
      \edef\UPM{\hexnumber\upmath@group}
      \new@mathgroup\amsa@group
      \define@mathgroup\mv@normal\amsa@group{msa}{m}{n}
      \define@mathgroup\mv@bold\amsa@group{msa}{m}{n}
      \edef\AMSa{\hexnumber\amsa@group}
      \makeatother
      \mathchardef\upi="0\UPM19
      \mathchardef\umu="0\UPM16
      \mathchardef\upartial="0\UPM40
      \mathchardef\leqslant="3\AMSa36
      \mathchardef\geqslant="3\AMSa3E
    \fi
  \fi
\fi 

\ifnfsstwo
  \DeclareMathAlphabet{\mathbfit}{OT1}{cmr}{bx}{it}
  \SetMathAlphabet\mathbfit{bold}{OT1}{cmr}{bx}{it}
  \DeclareMathAlphabet{\mathbfss}{OT1}{cmss}{bx}{n}
  \SetMathAlphabet\mathbfss{bold}{OT1}{cmss}{bx}{n}
  \ifAMStwofonts
    \ifCUPmtlplainloaded \else
      \DeclareSymbolFont{UPM}{U}{eur}{m}{n}
      \SetSymbolFont{UPM}{bold}{U}{eur}{b}{n}
      \DeclareSymbolFont{AMSa}{U}{msa}{m}{n}
      \DeclareMathSymbol{\upi}{0}{UPM}{"19}
      \DeclareMathSymbol{\umu}{0}{UPM}{"16}
      \DeclareMathSymbol{\upartial}{0}{UPM}{"40}
      \DeclareMathSymbol{\leqslant}{3}{AMSa}{"36}
      \DeclareMathSymbol{\geqslant}{3}{AMSa}{"3E}
    \fi
  \fi
\fi 

\ifCUPmtlplainloaded \else
  \ifAMStwofonts \else 
    \def\upi{\pi}
    \def\umu{\mu}
    \def\upartial{\partial}
  \fi
\fi

\title[Necessity of Dark Matter in Modified Newtonian Dynamics within Galactic Scales?
 - Testing the Covariant MOND in Elliptical Lenses]
 {Necessity of Dark Matter in Modified Newtonian Dynamics within Galactic Scales?
 - Testing the Covariant MOND in Elliptical Lenses}

\author[Chiu et al.]
       {M.C. Chiu$^1$, Y. Tian$^2$ and C.M. Ko$^3$ \\
        $^1$Institute of Astronomy, National Central University, Taiwan\\
        $^2$Department of Physics, National Central University, Taiwan\\
        $^3$Institute of Astronomy, Department of Physics and Center of Complex System,
        National Central University, Taiwan}
\date{}

\pagerange{\pageref{firstpage}--\pageref{lastpage}}
\pubyear{2008}

\begin{document}

\maketitle

\label{firstpage}

\begin{abstract}
Modified Newtonian Dynamics (MOND) and its relativistic version
$-$ TeVeS offer us an alternative perspective to understand the
universe without the demand of the elusive cold dark matter. This
MONDian paradigm is not only competitive with the conventional CDM
in a large range of scales, but also even more successful in the
galactic scale. Recently, by studying 6 lensing systems, Ferreras
et al. (2008) claimed that MOND still needs dark matter even in
galactic scales. When we study the same systems, however, we yield
an opposite conclusion. In this contribution, we report our result
and conclude that MOND does not need dark matter in galactic
lensing systems. Furthermore, we extend our study to 22 SLACS
(Sloan Lens ACS Survey) lenses, and obtain the same conclusion as
well, i.e., no dark matter is needed in elliptical galaxies.
\end{abstract}


\section{Introduction}\label{sec:intro}
%
%

Migrom's MOdified Newtonian Dynamics (MOND) is an alternative
to the conventional dark matter paradigm for the rotation
curve of galaxies and similar phenomena \cite{Mil83}.
TeVeS-a relativistic MOND theory \cite{Bek04} offered
an opportunity to explain relativistic phenomena such as
cosmology and gravitational lensing \cite{Chiu06}.

Recently, Ferreras et al. (2008) studied 6 lensing systems and
claimed that MOND still needs dark matter even in galactic scales.
We revisit the problem and arrive at an opposite conclusion. Here
we report our result on 10 lens from CASTLES (including the 6 lens
studied by Ferreras et al.) and 22 lens from SLACS.

\section{Lensing Equation}\label{sec:eq}
Since in the MONDian paradigm mass distribution only follows
baryon, we adopt the Hernquist model,
$|\nabla\Phi_{N}|=GM/(r+r_{h})^2$, to the lenses.
We incorporate our lensing formalism in a $\nu$HDM cosmological
background ($\Omega_{b}=0.05$, $\Omega_{\nu}=0.17$
$\Omega_{\Lambda}=0.78$, $h=0.7$).

In general, for any double-images lensing systems:
\begin{equation}\label{lensEq}
  \theta_{+}\theta_{-}={\theta_E}^2\left({{\theta_{+}f_{-}+\theta_{-}f_{+}}
  \over{\theta_{+}+\theta_{-}}}\right),
\end{equation}
where
$\theta_\pm$ are positions of the two images,
$\theta_E^2=4GMD_{\rm LS}/c^2D_{\rm L}D_{\rm S}$, and
$f_\pm$ are dimensionless functions, which depend on the mass model
of lenses and the forms of $\tilde{\mu}(x)$ in MOND.
We consider (i) the Bekenstein's form
$\tilde{\mu}=({-1+\sqrt{1+4x}})/({1+\sqrt{1+4x}})$ \cite{Bek04},
(ii) the simple form $\tilde{\mu}=x/(1+x)$ \cite{Fam05}, and
(iii) the standard form $\tilde{\mu}=x/\sqrt{1+x^2}$ \cite{San02}.
Here $x=a/a_0$, the ratio of the actual acceleration to the
acceleration parameter in MOND.

\begin{table}\label{castles}
\caption{Aperture mass (total mass) of 10 lenses from CASTLES
($10^{10} M_{\odot}$) in $\nu$HDM}
\begin{tabular}{lcrrr}
  Lens & Bekenstein & Simple & Standard \\
\hline
  Q$0142-100$    & 10.79 (18.36) & 13.66 (23.20) & 16.05 (27.31) \\
  HS$0818+1227$  & 18.14 (28.30) & 23.39 (36.17) & 27.79 (43.35) \\
  FBQ$0951+2635$ & 1.54 (2.16)  & 1.91  (2.67)  & 2.15 (3.01) \\
  BRI$0952-0115$ & 2.01 (2.48)  & 2.59  (3.21)  & 3.19 (3.93) \\
  Q$1017-207$    & 2.45 (5.89)  & 3.22  (7.55)  & 3.81 (9.15) \\
  HE$1104-1805$  & 45.17 (59.58) & 58.44 (77.08) & 71.78 (94.68) \\
  LBQ$1009-025$  & 7.71 (10.79)  & 9.76 (13.67)  & 11.53 (16.15) \\
  B$1030+071$    & 9.76 (16.61)  & 12.06 (20.51)  & 13.80 (23.47) \\
  SBS$1520+530$  & 11.91 (16.67)  & 15.20 (21.28)  & 18.08 (25.31) \\
  HE$2149-274$   & 7.04 (13.58)  & 8.96 (17.28)  & 10.67 (20.58) \\
\hline\hline
\end{tabular}


\begin{tabular}{lrrrrrrr}
Lens & Chabrier & Salpeter \\
\hline
  Q$0142-100$    & $ 20.9^{30.8}_{ 13.0}$  & $18.3^{32.2}_{13.2}$ \\
  HS$0818+1227$  & $ 16.2^{21.2}_{12.6}$   & $20.8^{28.1}_{13.4}$ \\
  FBQ$0951+2635$ & $ 1.1^{2.1}_{0.5}$      & $1.5^{3.0}_{0.8}$ \\
  BRI$0952-0115$ & $ 3.5^{4.0}_{2.7}$      & $4.4^{5.2}_{3.5}$ \\
  Q$1017-207$    & $ 4.3^{13.0}_{1.4}$     & $6.4^{19.0}_{2.3}$ \\
  HE$1104-1805$  & $ 22.8^{51.2}_{12.7}$   & $36.6^{63.7}_{23.1}$ \\
  LBQ$1009-025$  & $ 5.5^{7.9}_{4.2}$      & $ 7.4^{9.8}_{5.0}$  \\
  B$1030+071$    & $ 10.6^{15.3}_{6.5}$    & $ 14.5^{21.3}_{8.3}$ \\
  SBS$1520+530$  & $ 18.5^{30.9}_{11.2}$   & $21.8^{34.1}_{11.9}$ \\
  HE$2149-274$   & $ 4.6^{6.7}_{3.6}$      & $6.9^{8.9}_{5.0}$ \\
\end{tabular}
\end{table}

\begin{table}\label{slacs}
\caption{Mass of 22 lenses from SLACS ($10^{10} M_{\odot}$) in $\nu$HDM}
\begin{tabular}{lccc}
  Lens & $\left({\theta\over\theta_0}\right)_{\rm bek}$ &
         $\left({\theta\over\theta_0}\right)_{\rm spl}$ &
         $\left({\theta\over\theta_0}\right)_{\rm std}$ \\

\hline
  SDSS J$002907.8-005550$ & 0.227 & 0.200 & 0.183 \\
  SDSS J$015758.9-005626$ & 0.338 & 0.298 & 0.274 \\
  SDSS J$021652.5-081345$ & 0.152 & 0.135 & 0.126 \\
  SDSS J$025245.2+003958$ & 0.246 & 0.216 & 0.197 \\
  SDSS J$033012.1-002052$ & 0.304 & 0.267 & 0.243 \\
  SDSS J$072805.0+383526$ & 0.247 & 0.218 & 0.203 \\
  SDSS J$080858.8+470639$ & 0.253 & 0.224 & 0.205 \\
  SDSS J$090315.2+411609$ & 0.301 & 0.263 & 0.240 \\
  SDSS J$091205.3+002901$ & 0.113 & 0.101 & 0.095 \\
  SDSS J$095944.1+041017$ & 0.164 & 0.146 & 0.137 \\
  SDSS J$102332.3+423002$ & 0.242 & 0.214 & 0.199 \\
  SDSS J$110308.2+532228$ & 0.138 & 0.121 & 0.112 \\
  SDSS J$120540.4+491029$ & 0.181 & 0.161 & 0.150 \\
  SDSS J$125028.3+052349$ & 0.241 & 0.212 & 0.195 \\
  SDSS J$140228.1+632133$ & 0.189 & 0.167 & 0.156 \\
  SDSS J$142015.9+601915$ & 0.111 & 0.102 & 0.097 \\
  SDSS J$162746.5-005358$ & 0.181 & 0.160 & 0.149 \\
  SDSS J$163028.2+452036$ & 0.264 & 0.232 & 0.213 \\
  SDSS J$223840.2-075456$ & 0.178 & 0.158 & 0.148 \\
  SDSS J$230053.2+002238$ & 0.180 & 0.160 & 0.149 \\
  SDSS J$230321.7+142218$ & 0.163 & 0.145 & 0.135 \\
  SDSS J$234111.6+000019$ & 0.183 & 0.161 & 0.148 \\
\hline\hline
\end{tabular}


\begin{tabular}{lrrrr}
  Lens & $M_{\rm bek}$ & $M_{\rm spl}$ & $M_{\rm std}$ & $M_{\sigma}$ \\
\hline
  SDSS J$002907.8-005550$ & 15.22  & 19.68  & 23.45  & $36.8^{42.0}_{31.8}$ \\
  SDSS J$015758.9-005626$ & 29.68  & 38.12  & 45.13  & $77.9^{101.1}_{57.4}$ \\
  SDSS J$021652.5-081345$ & 119.91 & 151.50 &172.79  & $126.9^{142.8}_{111.9}$ \\
  SDSS J$025245.2+003958$ & 25.84  & 33.44  & 40.27  & $24.9^{28.2}_{21.8}$ \\
  SDSS J$033012.1-002052$ & 27.26  & 35.43  & 42.76  & $35.0^{41.3}_{29.1}$ \\
  SDSS J$072805.0+383526$ & 25.72  & 32.95  & 38.35  & $29.4^{32.0}_{26.7}$ \\
  SDSS J$080858.8+470639$ & 26.01  & 33.38  & 39.57  & $40.9^{44.3}_{37.6}$ \\
  SDSS J$090315.2+411609$ & 41.75  & 54.77  & 65.92  & $44.2^{54.1}_{35.2}$ \\
  SDSS J$091205.3+002901$ & 139.99 & 176.47 & 197.36 & $126.7^{135.0}_{118.6}$ \\
  SDSS J$095944.1+041017$ & 16.17  & 20.40  & 23.24  & $24.4^{27.3}_{21.6}$ \\
  SDSS J$102332.3+423002$ & 25.48  & 32.43  & 37.66  & $35.9^{39.9}_{32.1}$ \\
  SDSS J$110308.2+532228$ & 23.91  & 30.93  & 36.23  & $36.8^{40.8}_{32.9}$ \\
  SDSS J$120540.4+491029$ & 35.24  & 44.65  & 51.29  & $59.9^{64.8}_{55.1}$ \\
  SDSS J$125028.3+052349$ & 27.77  & 35.99  & 42.20  & $47.3^{52.0}_{42.7}$ \\
  SDSS J$140228.1+632133$ & 53.51  & 68.51  & 78.22  & $59.6^{66.5}_{54.2}$ \\
  SDSS J$142015.9+601915$ & 10.24  & 12.21  & 13.35  & $19.3^{26.9}_{12.7}$ \\
  SDSS J$162746.5-005358$ & 45.52  & 58.40  & 67.44  & $72.9^{79.2}_{66.8}$ \\
  SDSS J$163028.2+452036$ & 63.71  & 82.47  & 97.16  & $64.9^{71.6}_{58.4}$ \\
  SDSS J$223840.2-075456$ & 22.68  & 28.72  & 32.96  & $28.3^{31.1}_{25.6}$ \\
  SDSS J$230053.2+002238$ & 56.99  & 72.55  & 83.09  & $67.1^{74.5}_{60.1}$ \\
  SDSS J$230321.7+142218$ & 62.56  & 79.12  & 90.74  & $62.9^{70.0}_{56.1}$ \\
  SDSS J$234111.6+000019$ & 41.77  & 53.97  & 63.75  & $44.2^{49.2}_{39.5}$ \\
\end{tabular}
\end{table}


\section{Data and Result}\label{sec:data}
We apply 10 CASTLES lensing systems along with 22 SLACS
lenses to study strong lensing in TeVeS.

CASTLES Catalogue has the most complete list of strong lensing
systems. We examine 10 double-image lenses from the this
catalogue, of which the (aperture) stellar masses have been
estimated from stellar population synthesis with two initial mass
functions (IMF), Salpeter's and Chabrier's \cite{Fer05}.
In Table~\ref{castles} we list the aperture mass and total mass
computed from the three forms of $\tilde\mu(x)$
(Bekenstein, simple and standard).

We also work out the masses for 22 SLACS lens. SLACS lens come
from SDSS Luminous Red Galaxy and MAIN SDSS galaxy sample, in
which the dispersion velocity can be estimated. We use
the M-$\sigma$ relation in \cite{San00} to create a
M-$\sigma$-$\rm{R_{eff}}$ relation of V-band in MOND, and apply
this formula to compute the dynamical mass.
The comparison between lensing mass and dynamical mass is given in
Table~\ref{slacs}. $(\theta/\theta_0)$ is a measure of how close the
lensing system is to the deep MOND or the Newtonian regime. All the
22 systems are in the intermediate MOND regime, in which the form of
$\tilde\mu(x)$ is important. In Fig. 1, we compare the mass from
lensing and mass from velocity dispersion.


\section{Summary}\label{sec:summary}

Our investigation supports that there is no need of dark matter in elliptical galaxies.

Among the three choices of $\tilde{\mu}(x)$ of MOND, simple form seems to yield the
best fit with dynamical counterparts.
This echoes the dynamical studies in spiral galaxies \cite{Fam05,Fam07}.

\section*{Acknowledgment}
This work is supported in part by NSC96-2112-M-008-014-MY3.

\begin{figure}\label{Mass02}
\centerline{\epsfxsize=9cm\epsffile{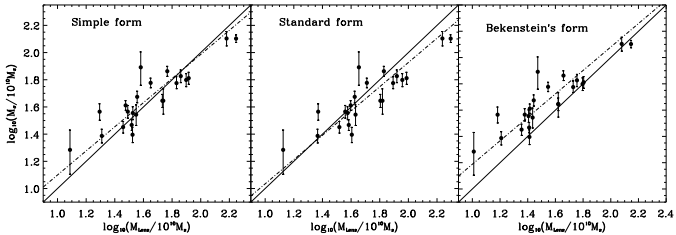}}
\caption[]{Comparison of mass from lensing and mass from dynamical measurement
(i.e., velocity dispersion). Simple form gives the best correlation.}
\end{figure}



\label{lastpage}

\clearpage

\end{document}